\RequirePackage[2020-02-02]{latexrelease}
\documentclass{rQUF2e}

\usepackage{epstopdf}
\usepackage{subfigure}

\theoremstyle{plain}

\theoremstyle{definition}

\theoremstyle{remark}

\begin{document}


\title{Variational Heteroscedastic Volatility Model}

\author{Zexuan Yin$^{\ast}$$\dag$\thanks{$^\ast$Corresponding author.
Email: zexuan.yin.20@ucl.ac.uk}, Danial Saef${\ddag}$ and Paolo Barucca${\dag}$\\
\affil{$\dag$Department of Computer Science, University College London, WC1E 7JE, United Kingdom\\
$\ddag$International Research Training Group, Humboldt Universit{\"a}t zu Berlin, Group 1792, Spandauer Str.1, 10178 Berlin, Germany\\
$\dagger$p.barucca@ucl.ac.uk, $\ddagger$danial.saef@hu-berlin.de} \received{v1.1 released April 2022} }

\maketitle

\begin{abstract}
We propose Variational Heteroscedastic Volatility Model (VHVM) - an end-to-end neural network architecture capable of modelling heteroscedastic behaviour in multivariate financial time series. VHVM leverages recent advances in several areas of deep learning, namely sequential modelling and representation learning, to model complex temporal dynamics between different asset returns. At its core, VHVM consists of a variational autoencoder to capture relationships between assets, and a recurrent neural network to model the time-evolution of these dependencies. The outputs of VHVM are time-varying conditional volatilities in the form of covariance matrices. We demonstrate the effectiveness of VHVM against existing methods such as Generalised AutoRegressive Conditional Heteroscedasticity (GARCH) and Stochastic Volatility (SV) models on a wide range of multivariate foreign currency (FX) datasets.     
\end{abstract}


\begin{classcode}C32, C45, C53, \end{classcode}

\section{Introduction}
Financial time series is known to exhibit heteroscedastic behaviour - time-varying conditional volatility \citep{Engle2007, Poon2003}. Being able to model and predict this behaviour is of practical importance to professionals in finance for the purpose of risk management \citep{Christoffersen2000, Long2020}, derivative pricing \citep{Duan1995}, and portfolio optimisation \citep{Rankovic2016, Escobar-Anel2022}. 

It is well documented in literature that (conditional) volatility is forecastable on hourly or daily frequencies \citep{Christoffersen2000}. On a univariate level, this involves predicting time-varying variances of asset returns; this predictability can be attributed to the so-called volatility clustering phenomenon: large (small) changes in asset price are often followed by further large(small) changes \citep{Engle2007, Fama1965, Schwert1989}. On a multivariate level (involving a portfolio of assets), volatility forecasting involves estimating conditional covariances between asset pairs in addition to conditional variances of the assets. One could argue that some of this predictability comes from the so-called spillover effect: the transfer of shock between different financial markets \citep{Jebran2017, Hassan2007, Du2011}. An effective multivariate volatility model therefore needs to capture both intra and inter-time series dynamics.

Traditional volatility forecasting models can be divided into two main categories: Generalised AutoRegressive Conditional Hertoscedasitcity (GARCH) \citep{Bollerslev1986, Nelson1991, GLOSTEN1993} and Stochastic Volatility (SV) \citep{Jacouier1994, Chan2016} models. The two competing classes of models rely on different underlying assumptions \citep{Luo2018}. GARCH models describe a deterministic relationship between future conditional volatility and past conditional volatility and squared returns. SV models assumes that conditional volatility follows a latent autoregressive process. Although there is no general consensus that GARCH is always superior to SV (or vice versa), there is some evidence that SV models are more flexible in  modelling the characteristics of asset returns \citep{Chan2016, Shapovalova2021}. Nonetheless, the popularity of GARCH models seems to surpass that of SV models due to several reasons. Firstly, GARCH models are easier to fit than SV models. The parameters of GARCH are obtained using maximum likelihood estimation, whereas for SV models one needs to obtain samples from an intractible posterior distribution using methods such as Markov chain Monte Carlo (MCMC), which works well when the number of parameters is small, however the convergence can be slow in larger models \citep{Shapovalova2021}. Secondly, there is an abundance of open-source software (packages) for GARCH models, such as fGarch \citep{fGarch} and rugarch \citep{rugarch} in the programming language R, and arch \citep{kevin_sheppard_2022_6400724} in Python. For SV models however, there was no go-to package for model estimation until the release of stochvol and factorstochvol \citep{Hosszejni2021} in R.

It is worth mentioning that GARCH models too suffer from the curse of dimensionality. For a portfolio of $n$ assets, the computational complexity of GARCH models scales with $O(n^5)$, which makes it impossible fit to beyond a portfolio of roughly 5 assets \citep{Wu2013}.

In recent years, the application of deep learning models for time series forecasting has achieved state of the art performances in many domains \citep{Bandara2019, Bose2017, Li2018}. Since our observations consist of multiple asset returns time series, a natural research direction would be to investigate whether deep learning models can capture complex dependencies between different assets across time. There are two main obstacles in this task. Firstly, the conditional volatility (covariance matrix) is a latent variable and must be inferred using observational data \citep{Luo2018}. Secondly, for a matrix to be a valid covariance matrix, it must be symmetric and positive definite \citep{Engle1995}. How to impose these constraints on a neural network such that its outputs are valid covariance matrices is a challenging task.     

To tackle the first challenge, we adopt a recent trend that combines a variational autoencoder (VAE) and a recurrent neural network (RNN) (a VRNN) to allow efficient structured inference over a sequence of continuous latent random variables \citep{Chung2015, Bayer2014, Krishnan2017, Fabius2015, Fraccaro2016, Karl2017}. The use of a VAE (and hence variational inference) translates posterior approximation into an optimisation task which can be solved using a neural network trained with stochastic gradient descent. The use of an RNN allows information from previous steps to be used in the modelling and forecasting of the latent variable in future steps. In fact, this promising framework has been explored in \cite{Yin2022} and \cite{Luo2018}. In \cite{Yin2022}, the authors proposed a neural network adaptation of the GARCH model which showed improved performance over traditional GARCH models. However, this model also suffers from the curse of dimensionality as it is still a GARCH model by design. Since the focus of this paper is on multivariate volatility, not being able to scale up to higher dimensions (beyond at least 5) is a big limitation. The authors in \cite{Luo2018} proposed a purely data driven approach to volatility forecasting under the VRNN framework which we used as a baseline model in our results section.  

To tackle the second challenge, one possible approach is to combine traditional econometrics models with deep learning models. Traditional econometrics models have well understood statistical properties and neural networks can be used to enhance the predictive power of the model. In \cite{Yin2022}, the authors use a neural network to parameterise the time-varying coefficients of the BEKK(1,1) model, which is a multivariate GARCH model proposed by \cite{Engle1995}. The BEKK model produces symmetric and positive definite covariance matrices by design and therefore no other constraints need to be applied to the neural network output. As mentioned previously however, many econometric models suffer from curse of dimensionality.  Instead, we follow the approach in \cite{Dorta2018} and design our neural network such that it outputs the Cholesky decomposition of the precision matrix (inverse of the covariance matrix) at every time step. We will show later on how this guarantees symmetry and positive definiteness. \citep{Engle1995}

Our main contribution is therefore an end-to-end neural network architecture capable of forecasting valid covariance matrices. We show in the results section that VHVM consistently outperforms existing GARCH and SV baselines on a range of multivariate FX portfolios.

The rest of the paper is structured as follows: in section \ref{sec: Related Work} we present a summary of the field of volatility forecasting and introduce some of the popular models; also discuss how they are related to our proposed model. In  section \ref{sec: Materials and Methods} we formally introduce VHVM: its generative and inference components, and model training and forecasting. In section \ref{sec: experiments} we outline the experiments and data used to show the effectiveness of VHVM against existing baselines from both traditional econometrics and the field of deep learning.

\section{Related Work}
\label{sec: Related Work}

For a financial asset with price $S_t$ at time $t$, its log returns are computed as $r_t = log(S_t)-log(S_{t-1})$. The returns process $r_t$ can be assumed to have a conditional mean $\mathbb{E}[r_t|\emph{I}_{t-1}] = 0$ and a conditional variance $\mathbb{E}[{r_t}^2|\emph{I}_{t-1}] = {\sigma_t}^2$, in other words $r_{t}|\emph{I}_{t-1} \sim \mathcal{N}(0,\sigma_{t}^2)$. The information set $\emph{I}_t$ describes all relevant available at time $t$: $\emph{I}_t=\{r_{1:t},{\sigma_{1:t}}^2\}$. The time variation of conditional variance $\sigma$ is known as heteroscedasticity and the aim of volatility forecasting is to model this behaviour \citep{Engle1995}. 
\subsection{Generalise Autoregressive Conditional Heteroscedasticity}
ARCH \citep{Engle1982} and GARCH \citep{Bollerslev1986} models have been dominating the field of volatility forecasting since the late 1900s due to their simple model form, explainability, and ease of estimation. Many GARCH model variants have since been proposed to account for well-known stylised facts about financial time series such as volatility clustering and leverage effect \citep{Engle2007}. The EGARCH \citep{Nelson1991} and GJR-GARCH \citep{GLOSTEN1993} models for example, were designed to specifically accommodate the leverage effect. The most general GARCH($p$,$q$) model proposed by \cite{Bollerslev1986} describes a deterministic relationship between future conditional volatility, past conditional volatility and squared returns:
\begin{equation}
	\sigma_{t}^2 = \omega + \sum_{i=1}^{p} \alpha_{i}r_{t-i}^2 + \sum_{j=1}^{q} \beta_{j}\sigma_{t-j}^2,
\end{equation}
where $p$ and $q$ are lag orders of the ARCH and GARCH terms, under which the returns process $r_t$ has an unconditional mean $\mathbb{E}[r_t] = 0$ and unconditional variance $\mathbb{E}[r_{t}^2]=\frac{\omega}{1-\alpha-\beta}$. 

In total, there are many hundreds of GARCH model variants, however there exists little consensus on when to use which GARCH models as their performances tend to vary with the nature and behaviour of the time series being modelled, for example, the leverage effect is frequently observed in stock returns but rarely seen in foreign exchange currency returns \citep{Engle2007}. In \cite{Hansen2005} the authors compared the performances of 330 GARCH variants on Deutsche Mark-US Dollar exchange rates and IBM returns and found that the GARCH(1,1) was not outperformed by any other model in the foreign exchange analysis. In the IBM stock returns analysis however, the authors found that GARCH(1,1) was inferior to models that explicitly accounted for the leverage effect. It has thus become common practice to explore various GARCH variants for the same task \citep{Chan2016, Chu2017, Malik2005}. 

For a portfolio of assets, models tend to be multivariate generalisations of the univariate GARCH model. In additional to modelling conditional variances for each asset, one also needs to model time varying covariances between different asset pairs. The output for a multivariate GARCH model is a time-varying covariance matrix which describes the instantaneous intra and inter-asset relationships. Notable examples of multivariate GARCH models include the VEC model \citep{Bollerslev1988}, the BEKK model \citep{Engle1995}, the GO-GARCH model \citep{VanDerWeide2002}, and the DCC-GARCH model \citep{Christodoulakis2002, Tse2002, Engle2002}. For our analysis, we used the DCC-GARCH (dynamic conditional correlation) model as a multivariate GARCH baseline. The key difference between DCC-GARCH and BEKK (a popular multivariate GARCH) is that BEKK assumes constant conditional correlation between assets, i.e. the change in the covariance between two assets with time is due to the changes in the two variances (but the conditional correlation is constant) \citep{Huang2010}. The constant conditional correlation (CCC) assumption is rather crude since during different market regimes one would expect the correlation between assets to vary. The DCC-GARCH is a generalisation of a CCC-GARCH that accounts for dynamic correlation. During the estimation procedure, various univariate GARCH models are fit for the assets, followed by estimations of the parameters for conditional correlation. 

When fitting a multivariate GARCH model under the assumption of normal innovations ($\boldsymbol{r}_{t} \sim \mathcal{N}(0,\boldsymbol{\Sigma}_{t})$), we seek to maximise the multivariate normal log likelihood function \citep{Bauwens2006}:

\begin{equation}
	\label{eqn: log normal}
	\mathcal{L}(\theta) = -\frac{1}{2}\sum_{t=1}^{T} (log|\boldsymbol{\Sigma}_{t}|+\boldsymbol{r}_{t}^T \boldsymbol{\Sigma}_{t}^{-1}\boldsymbol{r}_{t}),
\end{equation}
which becomes computationally expensive in higher dimensions since we are required (for a portfolio of $n$ assets) to invert an $n \times n$ covariance matrix $\boldsymbol{\Sigma}_{t}$ for every time step. One solution to alleviate this burden is to directly work with the precision matrix (inverse covariance matrix) instead: $\boldsymbol{P} = \boldsymbol{\Sigma}^{-1}$. Setting a neural network output to be a precision matrix rather than a covariance matrix allows us to compute the log likelihood straightaway; hence bypassing the expensive matrix inversion step during model training. When the actual covariance matrix is required during the testing phase, one could simply invert the precision matrix to obtain the covariance matrix \citep{Luo2018, Dorta2018}. 

\subsection{Stochastic Volatility}
Stochastic Volatility is an alternative class of models that rely on assumption that the log conditional variance follows a non-deterministic autoregressive AR($p$) (usually $p=1$) process \citep{Shapovalova2021}:
\begin{equation}
	\label{eqn: log volatility}
	ln\sigma_{t+1}^2 = \mu + \phi ln\sigma_{t}^2 + \sigma_{\eta}\eta_{t+1},
\end{equation}

where $\eta_t \sim N(0,1)$ describes the innovation of the log variance process. For the rest of the section we refer the log volatility $ln \sigma_t^2$ as $h_t$ such that $r_t = exp(h_t/2)\epsilon_t$ where $\epsilon_t \sim N(0,1)$. 

In a multivariate setting, we seek to simultaneously model the volatility movements of a group of assets \citep{Platanioti2005}. Related movements between different asset classes, financial markets or exchange rates are often observed due to them being infuenced by common unobserved drivers (or factors) \citep{Aydemir1998}. \cite{Diebold1989} investigated the behaviour of seven dollar exchange rates for a period of 12 years and found that the seven series showed similarities in volatility behaviour in response to actions taken by the US government such as intervention efforts. Since stochastic volatility models are defined in terms of the log volatility process, it is harder to generalise a univariate model to its multivariate counterpart than a GARCH model \citep{Platanioti2005}. 

In this paper we take as baseline the factor model independently proposed by \cite{Pitt1999} and \cite{Aguilar2000}. An open-source package (factorstochvol) was developed in the programming language R by \cite{Hosszejni2021} which we used to run the baseline SV model in our analysis. The factor volatility model \citep{Hosszejni2021} for a portfolio of $n$ assets assumes $m$ latent common factors where $m<n$. We have that:

\begin{equation}
	\label{eqn: 4}
	\begin{aligned}
	\boldsymbol{r}_{t}|\boldsymbol{f}_{t} \sim \mathcal{N}(\boldsymbol{\Lambda}\boldsymbol{f}_t,\bar{\boldsymbol{\Sigma}}_t), \\
	\boldsymbol{f}_{t} \sim \mathcal{N}(0,\check{\boldsymbol{\Sigma}}_t),
	\end{aligned}
\end{equation}

where $\boldsymbol{f}_{t} = (f_{t1},...,f_{tm})^T$ is the vector of $m$ factors, $\boldsymbol{\Lambda} \in \mathbf{R}^{n\times m}$ is a matrix of factor loadings. The covariance matrices $\bar{\boldsymbol{\Sigma}}_t$ and $\check{\boldsymbol{\Sigma}}_t$ are diagonal and are defined as:

\begin{equation}
	\begin{aligned}
		\bar{\boldsymbol{\Sigma}}_t = diag(exp(\bar{h}_{t1}),...,exp(\bar{h}_{tn})), \\
		\check{\boldsymbol{\Sigma}}_t = diag(exp(\check{h}_{t1}),...,exp(\check{h}_{tm})),
	\end{aligned}
\end{equation}

where $\bar{h}$ and $\check{h}$ are the log variances of the $n$ assets and $m$ latent factors defined as follows (the AR(1) process given in (\ref{eqn: log volatility})):

\begin{equation}
	\begin{aligned}
		\bar{h}_{ti} \sim \mathcal{N}(\bar{\mu}_{i}+\bar{\phi}_{i}(\bar{h}_{t-1,i}), \bar{\sigma}_{i}^2), i = 1,...,n, \\
		\check{h}_{tj} \sim \mathcal{N}(\check{\mu}_{j}+\bar{\phi}_{j}(\check{h}_{t-1,j}), \check{\sigma}_{j}^2), j = 1,...,m.
	\end{aligned}
\end{equation}

Given the above, the multivariate returns process follows a 0 mean multivariate normal distribution with $\boldsymbol{r}_{t} \sim \mathcal{N}(0,\boldsymbol{\Sigma}_{t})$, where $\boldsymbol{\Sigma}_{t} = \boldsymbol{\Lambda}\check{\boldsymbol{\Sigma}}_t\boldsymbol{\Lambda}^T + \bar{\boldsymbol{\Sigma}}_t$. We see that the factor volatility model in (\ref{eqn: 4}) is by nature a state space model with a random walk latent transition process and a linear emission process $\boldsymbol{r}_{t}|\boldsymbol{f}_{t}$. We will show later on that our end-to-end neural network architecture follows the same theoretical framework: a neural network (an RNN) that models the non-linear latent transition process $\boldsymbol{f}_{t}|\boldsymbol{f}_{t-1}$, a neural network (VAE) that infers the latent factors from observational data  $\boldsymbol{f}_{t}|\boldsymbol{r}_{1:t}$, and a neural work (multilayer perceptron (MLP)) that parameterises the emission distribution $\boldsymbol{r}_{t}|\boldsymbol{f}_{t}$.

\subsection{Amortised Variational Inference}
For a state space model with latent variable $\boldsymbol{z}_{t}$ and observations $\boldsymbol{r}_{t}$, we are interested in the posterior distribution $P(\boldsymbol{z}_{t}|\boldsymbol{r}_{1:t})$ - also known as the filtering distribution in time series literature. Note that in the previous section we denoted the latent factors of a SV model as $\boldsymbol{f}_{t}$ as this is a common choice of notation in SV literature. From this section onwards we will use $\boldsymbol{z}_{t}$ instead of $\boldsymbol{f}_{t}$ since $\boldsymbol{z}_{t}$ is more commonly used to represent latent variables in machine learning literature. 

The variational autoencoder (VAE) \citep{Kingma2014} is a neural network architecture trained with stochastic gradient descent. VAE consists of an encoder neural network that parameterises the posterior distribution $P(\boldsymbol{z}|\boldsymbol{r})$, and a decoder neural network that parameterises the emission distribution $P(\boldsymbol{r}|\boldsymbol{z})$. Here we have emitted the subscript $t$ since VAEs were traditionally designed to work in a static setting and has been used extensively as generative models in computer vision \citep{Dorta2018}. Our aim is maximise the marginal log likelihood $logP_{\theta}(\boldsymbol{r})$ where the latent variable $\boldsymbol{z}$ has been integrated out and $\theta$ represents the model parameters we are optimising over. This task however involves an intractable integral.

\begin{equation}
	logP_{\theta}(\boldsymbol{r}) = log \int P_{\theta}(\boldsymbol{r}|\boldsymbol{z})P_{\theta}(\boldsymbol{z})\,d\boldsymbol{z}.
\end{equation}

With variational inference \citep{Blei2017}, we use a much simpler distribution $q_{\phi}(\boldsymbol{z}|\boldsymbol{r})$ to approximate the actual intractable posterior $P_{\theta}(\boldsymbol{z}|\boldsymbol{r})$. We can express the marginal log likelihood in terms of an lower bound - evidence lower bound (or ELBO) - and the Kullback Leibler divergence between our variational approximation  $q_{\phi}(\boldsymbol{z}|\boldsymbol{r})$ and actual posterior $P_{\theta}(\boldsymbol{z}|\boldsymbol{r})$:

\begin{equation}
	\label{eqn: logp}
	logP_{\theta}(\boldsymbol{r}) = ELBO(\theta,\phi) + KL(q_{\phi}(\boldsymbol{z}|\boldsymbol{r})||P_{\theta}(\boldsymbol{z}|\boldsymbol{r}));
\end{equation}

since $logP_{\theta}(\boldsymbol{r})$ depends only on $\theta$, minimising $KL(q_{\phi}(\boldsymbol{z}|\boldsymbol{r})||P_{\theta}(\boldsymbol{z}|\boldsymbol{r}))$ with respect to $\phi$ is equivalent to maximising $ELBO(\theta,\phi)$ w.r.t. $\phi$. Mximising $ELBO(\theta,\phi)$ w.r.t. $\theta$ corresponds to maximising  $logP_{\theta}(\boldsymbol{r})$. 

In a variational autoencoder \citep{Kingma2014} the $ELBO$ is expressed as:

\begin{equation}
	ELBO(\theta, \phi) = \mathbb{E}_{\boldsymbol{z} \sim q_{\phi}(\boldsymbol{z}|\boldsymbol{r})} [logP_{\theta}(\boldsymbol{r}|\boldsymbol{z})] - KL(q_{\phi}(\boldsymbol{z}|\boldsymbol{r})||P_{\theta}(\boldsymbol{z})),
\end{equation}

where $|P_{\theta}(\boldsymbol{z})$ is the prior distribution over $\boldsymbol{z}$, which is usually set to an uninformative $\mathcal{N}(0,\mathbf{\emph{I}})$. We seek to maximise the $ELBO(\theta,\phi)$ using the encoder neural network $q_{\phi}(\boldsymbol{z}|\boldsymbol{r})$ and decoder neural network $P_{\theta}(\boldsymbol{z}|\boldsymbol{r})$:

\begin{equation}
	\{\theta^*,\phi^*\}=\operatorname*{argmax}_{\theta,\phi}ELBO(\theta,\phi). 
\end{equation}

We follow a recent trend which adapts a VAE from a static model to a sequential model \citep{Chung2015,Bayer2014,Krishnan2017,Fabius2015,Fraccaro2016,Karl2017}. This usually involves using another neural network to parameterise a learned prior distribution $P_{\theta}(\boldsymbol{z}_{t}|\boldsymbol{z}_{t-1})$ to replace the uninformative prior $\mathcal{N}(0,\mathbf{\emph{I}})$. This learned prior describes the latent transition dynamics of the state space model, similar to $\boldsymbol{f}_{t}$ of the stochastic volatility model in (\ref{eqn: 4}) but conditioned on $\boldsymbol{f}_{t-1}$ as opposed to a random walk. The decoder neural network parameterises $P_{\theta}(\boldsymbol{r}_t|\boldsymbol{z}_t)$, which corresponds to the emission mechanism $\boldsymbol{r}_{t}|\boldsymbol{f}_{t}$ in (\ref{eqn: 4}). The posterior approximated by the encoder netowrk becomes $P_{\theta}(\boldsymbol{z}_{t}|\boldsymbol{r}_{1:t})$ as opposed to $P_{\theta}(\boldsymbol{z}|\boldsymbol{r})$ in a static setting. This requires a sequence to serve as input into an inference model, which is carried by an RNN since the hidden state $\boldsymbol{h}_{t}$ is a summary of the sequence $\boldsymbol{r}_{1:t}$.

The VAE-RNN (or VRNN) framework has achieved state of the art performance in sequence modelling tasks such as video prediction \citep{Franceschi2020,Denton2018}; hence we also leveraged this framework to design our volatility model.

\section{Materials and Methods}
\label{sec: Materials and Methods}
\subsection{Covariance Matrix Parameterisation}
A covariance matrix is required to be both symmetric and positive definite \citep{Engle1995}. Under the assumption that the returns time series follows a multivariate normal distribution $\boldsymbol{r}_{t} \sim \mathcal{N}(0,\boldsymbol{\Sigma}_{t})$, we wish to evaluate the log determinant $log|\boldsymbol{\Sigma}_{t}|$ and Mahalanobis distance $\boldsymbol{r}_{t}^T \boldsymbol{\Sigma}_{t}^{-1}\boldsymbol{r}_{t}$ from the log likelihood (\ref{eqn: log normal}). We follow the parameterisation scheme in \cite{Dorta2018} and perform a Cholesky decomposition on the precision matrix:

\begin{equation}
	\boldsymbol{P}_t = \boldsymbol{\Sigma}_{t}^{-1} = \boldsymbol{L}_{t}\boldsymbol{L}_{t}^{T},
\end{equation}
which ensures $\boldsymbol{P}_t$ is symmetric by construction. To ensure positive definiteness, we require that the diagonal entries of $\boldsymbol{L}_{t}$ to be strictly positive; this could achieved by applying a Softplus function on top of the neural network output. We see from the log likelihood (\ref{eqn: log normal}) that the covariance matrix needs to be inverted before evaluating the Mahalanobis distance; this process is costly at higher dimensions. Working with the precision matrix allows us to bypass the inversion during model training as the Mahalanobis distance is simply $\boldsymbol{r}_{t}^T \boldsymbol{L}_{t}\boldsymbol{L}_{t}^{T}\boldsymbol{r}_{t}$. To evaluate the log determinant, we have  $log|\boldsymbol{\Sigma}_{t}| = -2\sum_{i=1}^{n}log(l_{ii,t})$, where $l_{ii,t}$ is the $i^{th}$ element in the diagonal of $ \boldsymbol{L}_{t}$. Under this scheme, for a portfolio of $n$ assets, the output of our neural network is simply a vector of size $\frac{n(n+1)}{2}$ (denoted $\boldsymbol{z}_t$ thereon). We convert the vector $\boldsymbol{z}_t$ into a lower triangular matrix in a deterministic way using the torch.tril\_indices() method in PyTorch (e.g. $f([a,b,c]^T) = \begin{bmatrix}
	a & 0 \\
	b & c 
\end{bmatrix}$). We then apply a Softplus function to the diagonal elements of this matrix (to ensure positive definiteness) and the resulting matrix is the lower Cholesky matrix $\boldsymbol{L}_{t}$. This procedure is carried out at every time step to produce time-varying precision matrices. When the actual covariance matrix is required, for example in the test set to evaluate model performance, matrix $\boldsymbol{P}_{t}$ is inverted to obtain $\boldsymbol{\Sigma}_{t}$

\subsection{Generative Model}
The generative model defines the joint distribution  $P_\theta(\boldsymbol{r}_{1:T}, \boldsymbol{L}_{1:T},\boldsymbol{z}_{1:T})$, where $\boldsymbol{r}_t$ is the multivariate returns process; $\boldsymbol{L}_t$ is the lower Cholesky decomposition of the precision matrix  $\boldsymbol{P}_t$; vector $\boldsymbol{z}_t$ is the neural network output of size  $\frac{n(n+1)}{2}$, which is the latent variable that we try to infer using observational data. We factorise the joint distribution as follows:

\begin{equation}
	\label{eqn: joint}
	P_\theta(\boldsymbol{r}_{1:T}, \boldsymbol{L}_{1:T},\boldsymbol{z}_{1:T}) = \prod_{t=1}^{T}P_\theta(\boldsymbol{r}_t|\boldsymbol{L}_t)P_\theta(\boldsymbol{L}_t|\boldsymbol{z}_t)P_\theta(\boldsymbol{z}_t|\boldsymbol{r}_{1:t-1}),
\end{equation}

where $P_\theta(\boldsymbol{z}_t|\boldsymbol{r}_{1:t-1})$ is a learned prior distribution which describes the transition dynamics of the latent variable $\boldsymbol{z}_t$. Information about the sequence $\boldsymbol{r}_{1:t-1}$ is carried by an RNN known as the gated recurrent unit (GRU) \citep{Cho2014} with hidden state $\boldsymbol{h}_{t}$ such that:

\begin{equation}
	P_\theta(\boldsymbol{z}_t|\boldsymbol{r}_{1:t-1}) = P_\theta(\boldsymbol{z}_t|\boldsymbol{h}_{t-1}) = \mathcal{N}(\boldsymbol{\mu}_{z,t},\boldsymbol{\Sigma}_{z,t}).  
\end{equation}

The prior distribution is parameterised by a multilayer perceptron (MLP):

\begin{equation}
	\{\boldsymbol{\mu}_{z,t},\boldsymbol{\Sigma}_{z,t}\}_{prior} = MLP_{Gen}(\boldsymbol{h}_{t-1}).
\end{equation}

\begin{figure}[h]
	\centering
	\includegraphics[width=0.75\columnwidth,clip,keepaspectratio]{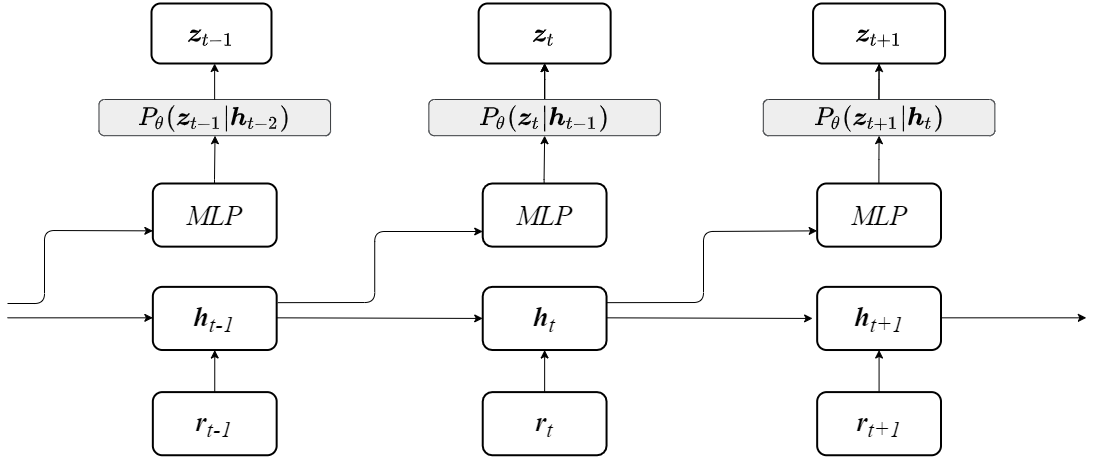}
	\caption{Generative model of VHVM. The generative MLP takes as input $\{\boldsymbol{r}_{1:t-1}\}$ and predicts the next-period latent factor $\boldsymbol{z}_{t}$, conditioned on which we can obtain an estimate of the covariance matrix $\boldsymbol{\Sigma}_{t}$.}
	\label{generative}
\end{figure}

$P_\theta(\boldsymbol{L}_t|\boldsymbol{z}_t)$ is a delta distribution centered on the output of the deterministic function: torch.tril\_indices followed by a Softplus function on the diagonal elements, which converts neural network output vector $\boldsymbol{z}_{t}$ into $\boldsymbol{L}_{t}$. The emission distribution $P_\theta(\boldsymbol{r}_t|\boldsymbol{L}_t)$ (the decoder) describes the 0 mean multivariate normal likelihood given in (\ref{eqn: log normal}) since $P_\theta(\boldsymbol{r}_t|\boldsymbol{L}_t) = P_\theta(\boldsymbol{r}_t|(\boldsymbol{L}_t\boldsymbol{L}_t^T)^{-1}=\boldsymbol{\Sigma}_t)$. A graphical presentation of the generative model is given in Fig \ref{generative}. We refer to the parameters of the generative model collectively as $\theta$, and the parameters of the inference model as $\phi$, which are jointly optimised using stochastic gradient variational Bayes.

There are various ways to design the prior distribution $P_\theta(\boldsymbol{z}_t|\emph{I}_{t-1})$, where $\emph{I}_{t-1} = \{\boldsymbol{r}_{1:t-1}, \boldsymbol{z}_{1:t-1}, \boldsymbol{\Sigma}_{1:t-1}\}$ represents all available information up to time $t-1$. We tested other design schemes such as $P_\theta(\boldsymbol{z}_t|\boldsymbol{r}_{1:t-1}, \boldsymbol{z}_{1:t-1})$ and $P_\theta(\boldsymbol{z}_t|\boldsymbol{r}_{1:t-1}, \boldsymbol{\Sigma}_{1:t-1})$, and found that in general the temporal dynamics of the latent variable could be well predicted using past returns alone; hence we decided on $P_\theta(\boldsymbol{z}_t|\boldsymbol{r}_{1:t-1})$. Choosing the prior this way keeps the number of neural network parameters lower than the other two specifications, which reduces overfitting; also we do not need to evaluate the covariance matrix during training.

\subsection{Inference Model}

The inference model defines the joint distribution $q_\phi(\boldsymbol{L}_{1:T},\boldsymbol{z}_{1:T}|\boldsymbol{r}_{1:T})$ which we factorise as follows:

\begin{equation}
	q_\phi(\boldsymbol{L}_{1:T},\boldsymbol{z}_{1:T}|\boldsymbol{r}_{1:T}) = \prod_{t=1}^{T}q_\phi(\boldsymbol{L}_t|\boldsymbol{z}_t)q_\phi(\boldsymbol{z}_t|\boldsymbol{r}_{1:t}),
\end{equation}

where the posterior distribution over latent variable $q_\phi(\boldsymbol{z}_t|\boldsymbol{r}_{1:t})$ is parameterised by the encoder (MLP) of the VAE:

\begin{equation}
	\{\boldsymbol{\mu}_{z,t},\boldsymbol{\Sigma}_{z,t}\}_{post} = MLP_{Inf}(\boldsymbol{h}_{t});
\end{equation}

this represents the filtering distribution, which is our inference of $\boldsymbol{z}_{t}$ given the most up-to-date observational data $\boldsymbol{r}_{1:t}$.  Since $q_\phi(\boldsymbol{z}_t|\boldsymbol{r}_{1:t})$ is our variational approximation of the actual posterior $P_\theta(\boldsymbol{z}_t|\boldsymbol{r}_{1:t})$, we see from (\ref{eqn: logp}) that maximising the $ELBO(\theta,\phi)$ w.r.t. $\phi$ is equivalent to minimising the KL divergence between the variational posterior and the actual posterior. The deterministic function to obtain $\boldsymbol{L}_t$ given $\boldsymbol{z}_t$ is the same as in the generative model: i.e. $q_\phi(\boldsymbol{L}_t|\boldsymbol{z}_t) = P_\theta(\boldsymbol{L}_t|\boldsymbol{z}_t)$, since this simply torch.tril\_indices() followed by Softplus. 

To summarise, VHVM is consisted of three neural networks: (1) an MLP ($MLP_{Gen}$) for the prior prediction model $	P_\theta(\boldsymbol{z}_t|\boldsymbol{r}_{1:t-1})$, also known as the decoder of the VAE which models the transition of the latent variable; (2) an MLP ($MLP_{Inf}$) for the variational posterior $q_\phi(\boldsymbol{z}_t|\boldsymbol{r}_{1:t})$, which is the encoder the VAE; (3) a GRU with hidden states $\boldsymbol{h}_t$ to carry sequential information about the multivariate returns process \{$\boldsymbol{r}_{1:T}$\} and is shared by the generative and inference models.

\begin{figure}[h]
	\centering
	\includegraphics[width=0.75\columnwidth,clip,keepaspectratio]{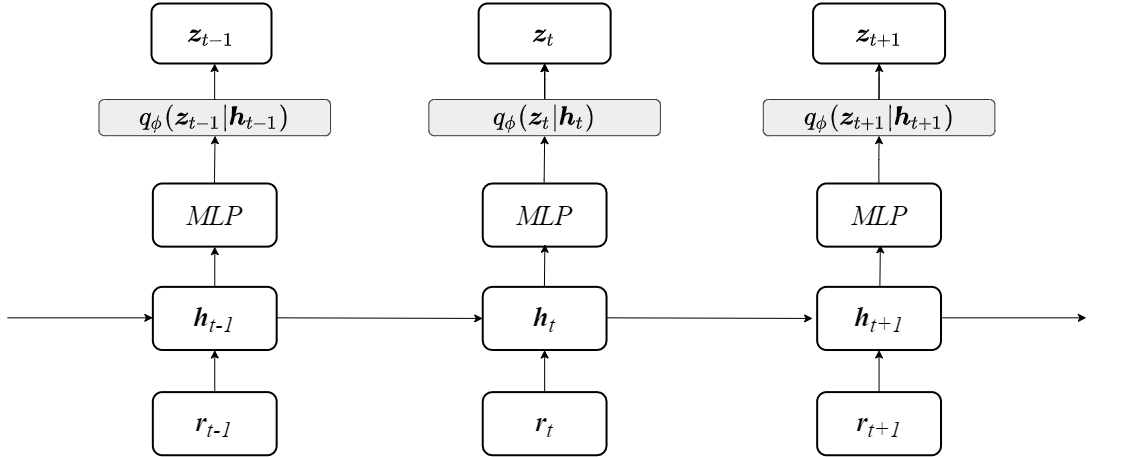}
	\caption{Inference model of VHVM. The Inference MLP takes as input $\{\boldsymbol{r}_{1:t}\}$ and paratermises the filter distribution over $\boldsymbol{z}_{t}$, conditioned on which we can obtain the posterior estimate of the covariance matrix $\boldsymbol{\Sigma}_{t}$.}
	\label{inference}
\end{figure}

\subsection{Model Training and Prediction}
To perform variational inference we seek to maximise the $ELBO(\theta,\phi)$ w.r.t. $\theta$ and $\phi$ jointly \citep{Kingma2014}. The expression for the evidence lower bound is given in (\ref{eqn ELBO}). VHVM is designed to output one step-ahead volatility prediction. When new observations become available, we update the hidden state $\boldsymbol{h}_t$ of the GRU, which serves as the input to the prediction network $MLP_{Gen}$ to predict the next period lower Cholesky matrix. 

\begin{equation}
	\label{eqn ELBO}
	ELBO(\theta, \phi) = \sum_{n=1}^{T}\mathbb{E}_{\boldsymbol{z}_{t} \sim q_\phi} [logP_\theta(\boldsymbol{r}_t|\boldsymbol{z}_t)] - KL(q_\phi(\boldsymbol{z}_t| \boldsymbol{r}_{1:t})||P_\theta(\boldsymbol{z}_t|\boldsymbol{r}_{1:t-1})), 
\end{equation}

\section{Experiments}
\label{sec: experiments}
We test VHVM on foreign exchange data obtained from the Trading Academy website (eatradingacademy.com). We compute daily log returns using data from the period 24/01/2012 to 23/01/2022 (a total of 3653 observations), from which we remove weekend readings where the change in asset price was 0. We constructed various portfolios using our collection of FX series for $n=$ 5, 10, 20, and 50. For model construction, we used a train:validation:test ratio of 80:10:10 and training for 50 epoches. 

For model benchmarking, we compared VHVM against three benchmarks: (1) the DCC-GARCH model \citep{Engle2002}, (2) a factor SV model with MCMC sampling \citep{Hosszejni2021}, and (3) Neural Stochastic Volatility model \citep{Luo2018}. The three baselines are representative models from the current approaches to volatility forecasting: GARCH models, SV models, and deep learning based models. We have chosen DCC-GARCH due to its ability to model dynamic conditional correlation between assets; we implemented the model in R using the package "rmgarch" \citep{Galanos2022}. For the factor SV model with MCMC sampler (MCMC-SV), we used the recently developed "factorstochvol" package in R \citep{Hosszejni2021}. 

The Neural Stochastic Volatility model (NSVM) \cite{Luo2018} is perhaps most relevant to our work since it was also designed under the VRNN framework. NSVM uses four recurrent neural networks to model temporal dynamics: one for the observed returns series $\boldsymbol{r}_{t}$ and another for the latent factor $\boldsymbol{z}_{t}$ in the generative model; similarly for the inference model. In our model however, we attempted to keep the number of model parameters low by using only one RNN but inputting the hidden state at different time steps to perform prediction and inference. Another key difference between NSVM and VHVM is that the output of NSVM is a low-rank approximation of the time-varying covariance matrix, whereas VHVM outputs the full covariance matrix. A low rank approximation may offer faster computations for higher dimensional portfolios, however we show that VHVM is consistently better in terms of performance.

For model evaluation, we perform one step-ahead covariance matrix forecasting on the test set, and following \cite{Wu2013} and \cite{Luo2018} use the log likelihood (\ref{eqn: log normal}) as our performance metric since it describes the likelihood of the observed data falling under our estimated distribution. We have also included the hyperparameters of our model and the baselines in the Appendix for reproduction purposes.

\section{Results and Discussion} 

In Table \ref{EUR} to \ref{MIX} we show the performance of VHVM against the three baseline models: NSVM, DCC-GARCH, and MCMC-SV on various 5 dimensional FX portfolios. For every time step we forecast a $5\times5$ covariance matrix and in the tables we report the cumulative log likelihood of the test set. We have highlighted in bold the best performing model in terms of log likelihood (higher is better). We observe that VHVM performs best in 17 out of the 20 constructed portfolios. The neural network baseline NSVM however performs best in only one of the portfolios. As previously mentioned, the two key difference between VHVM and NSVM are: (1) VHVM uses a single RNN to carry information about $\boldsymbol{r}_{1:t}$ and the hidden state at different time steps is used for forecasting ($\boldsymbol{h}_{t-1}$)/inference ($\boldsymbol{h}_{t}$), whereas NSVM uses four separate RNNs to model $\boldsymbol{z}_{t}$ and $\boldsymbol{r}_{t}$ in generation and inference; (2) NSVM outputs low rank approximations of the covariance matrix whereas VHVM outputs estimates of the full covariance matrix. We believe the simpler structure (fewer parameters) of our model has helped to reduce overfitting, and parameterising the full covariance matrix is more expressive than a low rank approximation at the expense of computational complexity ($O(n^2)$ vs $O(n)$)

\begin{table}[h]
	\centering
	\caption{Log likelihoods of 5 dimensional Euro-denominated portfolios. The best performing model is highlighted in bold; higher log likelihood is better. }
	\resizebox{0.8\columnwidth}{!}{
		\begin{tabular}{|c|c|c|c|c|}
			\hline
			FX pairs & VHVM (ours) & NSVM & DCC-GARCH & MCMC-SV\\
			\hline
			EURAUD, EURHKD, EURCAD, EURCNY, EURDKK & $\boldsymbol{-1013.489}$ & -1362.564 & -1134.183 & -1144.132\\
			\hline
			EURCNY, EURGBP, EURHKD, EURHUF, EURIDR & $\boldsymbol{-1189.944}$ & -1456.841 & -1235.496 & -1279.841\\
			\hline
			EURGBP, EURJPY, EURKRW, EURMXN, EURNOK & $\boldsymbol{-1418.791}$ & -1493.457 & -1506.990 & -1471.722\\
			\hline
			EURJPY, EURNZD, EURRUB, EURSGD, EURTHUB & $\boldsymbol{-1222.507}$ & -1331.243 & -1301.942 & -1262.875\\
			\hline
	\end{tabular}}
	\label{EUR}
\end{table}

\begin{table}[h]
	\centering
	\caption{Log likelihoods of 5 dimensional GBP-denominated portfolios. The best performing model is highlighted in bold; higher log likelihood is better. }
	\resizebox{0.8\columnwidth}{!}{
		\begin{tabular}{|c|c|c|c|c|}
			\hline
			FX pairs & VHVM & NSVM & DCC-GARCH & MCMC-SV\\
			\hline
			GBPAUD, GBPBGN, GBPBRL, GBPCAD, GBPCHF & $\boldsymbol{-1156.903}$ & -1564.931 & -1328.515 & -1300.182\\
			\hline
			GBPCHF, GBPCNY, GBPDKK, GBPHKD, GBPILS & $\boldsymbol{-898.588}$ & -1571.065 & -1047.304 & -1076.312\\
			\hline
			GBPCNY, GBPINR, GBPJPY, GBPMXN, GBPKRW & $\boldsymbol{-1142.915}$ & -1400.454 & -1248.320 & -1211.203\\
			\hline
			GBPRUB, GBPSEK, GBPTRY, GBPJPY, GBPCAD & -1639.355 & $\boldsymbol{-1628.892}$ & -2969.575 & -2900.870\\
			\hline
	\end{tabular}}
	\label{GBP}
\end{table}

\begin{table}[h]
	\centering
	\caption{Log likelihoods of 5 dimensional USD-denominated portfolios. The best performing model is highlighted in bold; higher log likelihood is better. }
	\resizebox{0.8\columnwidth}{!}{
		\begin{tabular}{|c|c|c|c|c|}
			\hline
			FX pairs & VHVM & NSVM & DCC-GARCH & MCMC-SV\\
			\hline
			USDAUD, USDBGN, USDCAD, USDCHF, USDCNY & $\boldsymbol{-1309.623}$ & -1663.140 & -1491.194 & -1333.658\\
			\hline
			USDCNY, USDEUR, USDGBP, USDHKD, USDNZD & -1416.474 & -1621.547 & -1536.093 & $\boldsymbol{-1414.045}$\\
			\hline
			USDEUR, USDHUF, USDINR, USDJPY, USDNZD & $\boldsymbol{-1237.078}$ & -1461.547 & -1306.365 & -1293.873\\
			\hline
			USDGBP, USDJPY, USDKRW, USDMXN, USDTRY & $\boldsymbol{-1609.177}$ & 1807.927  & -3490.222 & -3129.284\\
			\hline
	\end{tabular}}
	\label{USD}
\end{table}

\begin{figure}[h!]
	\centering
	\includegraphics[width=0.75\columnwidth,clip,keepaspectratio]{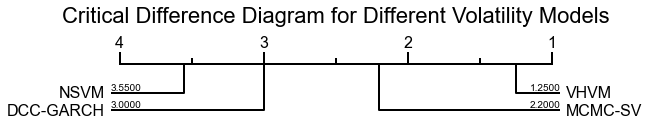}
	\caption{Critical difference diagram showing the comparison between VHVM, NSVM, DCC-GARCH, and MCMC-SV in 5 dimensional FX portfolios.}
	\label{cd-5d}
\end{figure}

To better gauge the relative performances of the four models, we follow \cite{IsmailFawaz2019} and plot a critical difference (CD) diagram showing the average ranking of the four model in Figure \ref{cd-5d}. Within a CD diagram, two models without a statistically significant difference (s.s.d.) in average ranking are connected with a horizontal line; the absence of such lines in Figure \ref{cd-5d} indicates that the four models are s.s.d. in performance across the 20 5 dimensional experiments. According to Figure \ref{cd-5d} VHVM has the best overall average ranking (1.25), followed by MCMC-SV(2.2), DCC-GARCH(3), and NSVM(3.55). The fact that MCMC-SV performs slightly better than DCC-GARCH is in accordance with claims that SV models are more flexible at modelling heteroscedastic behaviour in financial time series \citep{Shapovalova2021}.

\begin{table}[h]
	\centering
	\caption{Log likelihoods of 5 dimensional CNY-denominated portfolios. The best performing model is highlighted in bold; higher log likelihood is better. }
	\resizebox{0.8\columnwidth}{!}{
		\begin{tabular}{|c|c|c|c|c|}
			\hline
			FX pairs & VHVM & NSVM & DCC-GARCH & MCMC-SV\\
			\hline
			CNYCAD, CNYEUR, CNYGBP, CNYIDR, CNYJPY & $\boldsymbol{-1171.234}$ & -1447.516 & -1304.421 & -1272.866\\
			\hline
			CNYKRW, CNYMXN, CNYMYR, CNYRUB, CNYSEK & $\boldsymbol{-1270.803}$ & -1346.751 & -1384.705 & -1335.663\\
			\hline
			CNYGBP, CNYJPY, CNYSEK, CNYSGD, CNYTHB & $\boldsymbol{-1236.308}$ & -1427.729 & -1307.665 & -1283.031\\
			\hline
			CNYEUR, CNYMXN, CNYCAD, CNYUSD, CNYTHB & -1604.544 & -1592.762  & $\boldsymbol{-1535.586}$ & -1541.026\\
			\hline
	\end{tabular}}
	\label{CNY}
\end{table}

\begin{table}[h]
	\centering
	\caption{Log likelihoods of 5 dimensional mixed currency portfolios. The best performing model is highlighted in bold; higher log likelihood is better. }
	\resizebox{0.8\columnwidth}{!}{
		\begin{tabular}{|c|c|c|c|c|}
			\hline
			FX pairs & VHVM & NSVM & DCC-GARCH & MCMC-SV\\
			\hline
			EURAUD, GBPCAD, USDCHF, USDCNY, CNYGBP & $\boldsymbol{-1354.807}$ & -1580.274 & --1425.861 & -1363.748\\
			\hline
			EURHKD, GBPJPY, USDCHF, CNYRUB, CNYCAD & $\boldsymbol{-1139.711}$ & -1331.203 & -1290.668 & -1271.675\\
			\hline
			USDGBP, USDJPY, GBPCHF, CNYSGD, GBPMXN & $\boldsymbol{-1312.835}$ & -1379.446 & -1379.766 & -1313.195\\
			\hline
			CNYEUR, CNYGBP, EURKRW, USDINR, GBPRUB & $\boldsymbol{-1041.155}$ & -1247.187  & -1172.433 & -1146.655\\
			\hline
	\end{tabular}}
	\label{MIX}
\end{table}

In Table \ref{highD} we show the experimental results for larger portfolios (10, 20, and 50 dimensions). In these experiments we only compare VHVM and NSVM since both DCC-GARCH and MCMC-SV have difficulties scaling up to higher dimensions. The list of currencies included in each portfolio is included in the Appendix. We observe that VHVM performs better than NSVM across all higher dimensional portfolios, consistent with what we observe in the 5 dimensional experiments.

\begin{table}[h]
	\centering
	\caption{Log likelihoods of higher dimensional currency portfolios. The best performing model is highlighted in bold; higher log likelihood is better.}
		\begin{tabular}{|c|c|c|}
			\hline
			FX pairs & VHVM (ours) & NSVM \\
			\hline
			$10D_1$ & $\boldsymbol{-1302.786}$ & -3038.413\\
			\hline
			$10D_2$ & $\boldsymbol{-1804.729}$ & -3117.547\\
			\hline
			$10D_3$ & $\boldsymbol{-1472.019}$ & -3114.550\\
			\hline
			$10D_4$ & $\boldsymbol{-2831.592}$ & -3174.692\\
			\hline
			$20D_1$ & $\boldsymbol{789.993}$ & -6061.243\\
			\hline
			$20D_2$ & $\boldsymbol{-1334.264}$ & -6312.093\\
			\hline
			$50D$ & $\boldsymbol{-2073.592}$ & -15180.516\\
			\hline
	\end{tabular}
	\label{highD}
\end{table}

\section{Conclusion}
In this paper we propose Variational Heteroscedastic Volatility model (VHVM): an end-to-end neural network architecture capable of forecasting one step-ahead covariance matrices. VHVM outputs the lower Cholesky decomposition of a time-varying conditional precision matrix, which enforces two necessary constraints of a covariance matrix: symmetry and postive definiteness. Furthermore, by setting the neural network to output the precision matrix, we bypass the computationally expensive matrix conversion step in the evaluation of the multivariate normal log likelihood function. We demonstrated the effectiveness of VHVM against GARCH, SV, and deep learning baseline models and we observed that VHVM consistently outperformed its competitors. 

\section*{Acknowledgement}{The authors would like to thank Fabio Caccioli, Department of Computer Science, University College London, for proofreading the manuscript and providing feedback.}

\bibliographystyle{rQUF}
\bibliography{sample}
\bigskip
\medskip
\end{document}